# Structured epitaxial graphene: growth and properties

Yike Hu[1], Ming Ruan[1], Zelei Guo[1], Rui Dong[1], James Palmer[1], John Hankinson[1], Claire Berger[1,2], Walt A. de Heer[1]

[1]School of Physics, Georgia Institute of Technology, Atlanta GA 30332
[2] CNRS- Institut Néel, Grenoble, 38042, France

**Abstract**

Graphene is generally considered to be a strong candidate to succeed silicon as an electronic material. However, to date, it actually has not yet demonstrated capabilities that exceed standard semiconducting materials. Currently demonstrated viable graphene devices are essentially limited to micron size ultrahigh frequency analog field effect transistors and quantum Hall effect devices for metrology. Nanoscopically patterned graphene tends to have disordered edges that severely reduce mobilities thereby obviating its advantage over other materials. Here we show that graphene grown on structured silicon carbide surfaces overcomes the edge roughness and promises to provide an inroad into nanoscale patterning of graphene. We show that high quality ribbons and rings can be made using this technique. We also report on progress towards high mobility graphene monolayers on silicon carbide for device applications.

**Introduction**

Whether graphene will or will not become an important electronic material critically hinges on our ability produce high quality graphene and to pattern it at the nanoscale. Currently the predominant paradigm for producing graphene devices of any kind, is to start with an extended graphene sheet, that may be produced by a variety of methods[1-3], and subsequently to pattern the sheet using standard microelectronics lithography methods[4-8]. Typically lithography is performed by coating the graphene sheet with a resist material which is subsequently exposed to light or and electron beam, producing an image on the resist, that is subsequently developed to expose the graphene surface. The exposed graphene is then cut, typically using an oxygen plasma, that removes the exposed graphene resulting in a patterned graphene layer. However if patterned graphene is seen as a macro-molecule, then this "top down" method of producing it is very different from the method that is usually used to produce large molecules, either in the laboratory or in nature. In fact, the chemical approach would be to produce large molecules by assembling them from smaller ones, i.e. a "bottom up" approach[9].

While lithographically patterned structures are adequate for large devices, that rely primarily on the two-dimensional electronic properties of graphene, in fact, graphene's promise for electronics, as well as for new fundamental physics, relies primarily on its lower dimensional properties (i.e. nanoribbons and quantum dots). That is, graphene is most interesting when feature sizes are of the order of its Fermi wavelength so that quantum confinement effects manifest themselves. The carbon nanotube, with its well-known size and helicity dependent electronic properties is the archetype for quantum confined graphitic structures. In fact considerations of nanotube properties were the primary motivation for graphitic electronics research[1].

Graphene is a zero bandgap semiconductor (i.e. a semimetal). Modern digital microelectronics relies on the bandgap of the semiconductor so that field effect transistors can be produced with large on-to-off switching ratios in order to minimize leakage currents when the transistor is in the "off" state. However, this off condition cannot be reached with 2 dimensional graphene, not even at extremely low temperatures.

While devices based on two-dimensional graphene (i.e. devices with feature sizes greater than a 100 nm) are not considered to be useful for digital electronics, this material is useful for ultrahigh high frequency analog electronics[10]. Fast graphene transistors have been produced using 2 dimensional graphene. Even though these devices have very small on-to-off ratios (of the order of 10) they nevertheless have above unity gain at frequencies exceeding 200 GHz [11, 12], which brings them in the range of state of the art semiconducting devices[13]. In order to push these limits ever closer to the THz ranges will require further significant improvements of the graphene performance (as well as reductions in contact resistances and higher quality dielectrics). It should be mentioned that 2 dimensional graphene Hall bars patterned on epitaxial graphene are considered to be important in metrology as quantum Hall based resistance standards[14].

It was theoretically known that quantum confined structures like graphene ribbons, could have band gaps under certain conditions[15, 16]. The size of the bandgap $E_g$ is predicted to be of the order of $E_g$= 1 eV/W where W is the width of the ribbon. Consequently, theoretically, very narrow ribbons can have significant band gaps. There are however caveats. The size of the bandgap depends on the condition of the edge. In fact, theoretically, only graphene ribbons with armchair edges are semiconductors, and for them, only those where the width is an integer multiple of three hexagons[15, 16]. All other graphene ribbons, including those with zigzag edges and chiral ribbons that are intermediate between zigzag and armchair are all expected to be metallic (in general)[17]. Note, however, that theory does predict that bandgaps may open under certain circumstances[18].

It should therefore considered to be surprising that several recent papers report that lithographically patterned graphene nanoribbons from exfoliated graphene deposited on $SiO_2$ are all semiconducting with band gaps that are of the order of 1eV/$E_g$[5-8, 19]. The band gap was established by adjusting the charge density of the patterned graphene ribbon by applying a potential to a Si back gate in the usual way.

It was noted that with decreasing charge density the conductance dramatically decreases (by several orders of magnitude) similar to the conductance decrease that is observed in a semiconductor, from which the band gap was determined.

However, it was later determined[6-8, 19] that (in most cases) the band gap was actually a mobility gap. That is, the edges of these patterned ribbons is disordered causing scattering. The rough edges caused the graphene ribbon to resemble a series of "quantum dots" causing strong localization of the charge carriers and a large reduction in the mobility at low charge densities. While an interesting phenomenon in itself, materials that rely on such mobility gaps are not considered to be suitable for high-speed digital electronics.

In order to reduce edge scattering in narrow ribbons requires that edge roughness be reduced. This suggests significant modifications in the patterning method itself. Here we show examples of quality nano-structures produced directly on tailored silicon carbide surfaces without post-growth lithography processing. Progress towards high mobility epitaxial graphene monolayers for device applications is also reported.

**Structured monolayer graphene growth on silicon carbide**

Epitaxial graphene on silicon carbide ranks among the most promising candidates for large-scale graphene electronics. Epitaxial graphene grown on silicon carbide has several advantages over epitaxial graphene grown on metal substrates. One advantage is, that the material does not need to be transferred from the metal to another dielectric substrate. The SiC substrate is a large band-gap semiconductor widely used in the industry[20] and schemes are developed for integration with Silicon[21]. The interface between epitaxial graphene and the silicon carbide substrate is reasonably well understood on the atomic scale[22]. One of the most important advantages of epitaxially grown graphene sheets, is that here are no trapped impurities under the graphene (although interface can be modified by passivation and intercallation), and it can be readily patterned. Large-area graphene (on both polar faces of SiC) has been grown by several groups [1, 12, 14, 21, 23-29]. There are still progress to be made, however, before large area graphene layers with uniform properties are produced.

Another advantage is, that graphene growth can be directed. That is, by proper tailoring of the substrate, graphene can be grown only where it is desired as shown in detail below. This method of graphene growth circumvents the damaging lithographic patterning step that causes the large mobility drops in the patterned graphene ribbons mentioned above. Moreover, surprisingly, all graphene ribbons produced by this method are found to be metallic. There is no evidence for a band gap, in contrast to lithographically patterned graphenes that show evidence of band gaps or mobility gaps. These observation indicate that scattering at the edges is suppressed compared with ribbons produced using standard lithography methods.

Structured graphene growth is pursued along two distinct tracts: large area graphene growth on the carbon terminated surface, and sidewall graphene structured growth[30] on the silicon terminated surfaces. Both of these approaches are treated in detail below.

It is worth mentioning, that the often-stated disadvantage of epitaxial graphene on silicon carbide substrates, is the price of the substrate itself. This would indeed be an important factor if the aim were to compete with silicon based electronics. However, the ultimate goal is to succeed silicon, that is to build devices that exceed silicon in at least one key parameter, be it feature size, speed or power consumption. In those cases the substrate cost (which even now is minor compared to the processing cost) is not important. There are many examples of disruptive technologies that were initially costly but significantly outperformed prior technologies: semiconductor electronics and air travel are prime examples.

**Structured C-face graphene growth**

It is well known that the mobilities of graphene monolayers grown on the (000-1) face (the C-face) are significantly greater (typically by a factor of 3-10) than those graphene on the (0001) face (the Si-face). For this reason considerable effort has expended to control the graphene growth on the C-face, in particular for the 2-dimensional mono graphene devices mentioned above. High quality graphene mono- and multilayers have are grown on the C-face of hexagonal silicon carbide (either 6H or 4H) using the confinement controlled sublimation method (CCS)[23]. Briefly, in this method, a silicon carbide chip is heated in vacuum, inside a graphite capsule, that is supplied with a small leak. One or more graphene layers form at temperatures between 1500-1600 C. Uniform monolayer graphene growth is difficult to achieve over the entire 4 x 5 mm SiC chip surface. In particular, while large area growth has been accomplished consisting of a continuous graphene layer covering the entire surface, below this layer (which over most of the area is a monolayer) there are occasionally patches with additional graphene layers. Even though the top-most layer is continuous, these patches affect the transport properties, and ideally they should be eliminated. This sub-layer partial growth is more critical on the Si-face because A-B stacked bilayers don't have the same electronic structure as monolayers. On the C-face the monolayer electronic structure is preserved for multilayers due to the orientational stacking (see below)[31, 32] For quantum-Hall effect-based metrology applications it is important that the graphene is uniform and continuous on predetermined locations, that can then be patterned. While high quality continuous and uniform multilayer epitaxial graphene films are relatively easy to produce[23], uniform monolayers are still challenging. Progress is described below.

Graphene growth on the C-face nucleates primarily at natural substrate steps [25] (caused by step bunching), etched steps, and defects like screw dislocations (Figure 1). The monolayer growth patches at steps (Fig. 1a) are uniformly distributed over the silicon carbide chip surface. However, growth at screw dislocations (Fig. 1b) is not controlled and typically results in rather deep cervices covered with mulilayer graphene. It is clear that screw dislocations are an impediment to uniform growth.

Raman spectroscopy is a powerful characterization tool for graphene deposited on silicon oxide and it has been extensively treated in the literature[33]. Raman spectra for epitaxial graphene is basically similar however with some significant and important differences[26, 32]. Most importantly, the Raman spectrum of the silicon carbide substrate itself is superimposed on the graphene spectrum with intense features in the range 1500 to 2000 cm$^{-1}$ (Fig. 2a, red curve). The other characteristic bands (D, G and 2D) are used for diagnostic purposes. For Bernal stacked multilayer graphene, the 2D band evolves as a superposition of peaks due to the evolution of the electronic structure with increasing number of layers that converges to the graphite bandstructure[33]. However the evolution in C-face graphene is subtler since each layers has the same electronic structure as that of a monolayer . Consequently, the Raman spectra of thick C-face graphene multilayers are essentially identical to that of a monolayer[32]. However, the 2D band does broaden for the first few layers. This broadening is (most likely) primarily caused by the charge density variations from one layer to the next : the bottom most layer is negatively charged (n~5 10$^{-12}$ cm$^{-2}$) due to the SiC substrate, while the others are essentially neutral. Shifts in the 2D band due to change in the Fermi level have been discussed in ref [34]. Moreover, the top-most layer is (often) *p* doped due to environmental effects. In any case, as for exfoliated graphene on silicon carbide, monolayer graphene on SiC can be identified by its signature 2D band.

The positions of the G band and the 2D band as a function of charge density for monolayer graphene (on the C-face) are shown in Fig. 2. The charge densities have been determined from the Hall resistance. Note that the charge on a graphene monolayer is the sum of the interface charge and the environmentally induced charge. An epitaxial graphene monolayer is intrinsically negatively charged but it gradually becomes positively charged after exposure to the ambient. Heating in vacuum restores its virgin condition. An increase in the energies of these two Raman bands with increasing charge density (either positive or negative) is also observed in exfoliated graphene. While the increase in the G band agrees quantitatively with that observed for electrostatically gated exfoliated graphene, the shift in the 2D band is significantly larger than that observed in exfoliated graphene. We further note that in contrast to exfoliated graphene, the ratio of the 2D peak intensity and the G band intensity is not a reliable indicator for monolayer graphene as it is found vary from 2 to 9. However, we have found that ratio of the intensity of the G band and the SiC bands is a reliable indicator of the graphene layer thickness, that is particularly accurate in identifying graphene monolayers on the C face. The D band is typically very small in C face graphene indicating low defect densities in the material.

In Figure 3 we present a more detailed analysis of a C-face monolayer including both topographic AFM data and scanning Raman spectroscopy data. Usually a epitaxial graphene on the C-face can be recognized by the pleats as shown in Fig. 3a, however, occasionally an anomalous pleat-free region is observed within a monolayer patch. A micro Raman map of this patch (Fig. 3 b-c) shows that the areas with pleats have a normal Raman spectrum with a peak 2D intensity at 2680 cm$^{-1}$, while the Raman spectrum in the anomalous region has its 2D peak intensity at 2760 cm$^{-1}$. This shift in the Raman band indicates that this region is probably stressed and more strongly coupled to the substrate than normal.

Attempts to structure silicon carbide, by etching mesas or pits, in order to direct graphene growth on the C face have had good success (Fig.4). It appears that graphene growth initiates at specific corners of the hexagonal pit etched in C-face SiC. Further work will reveal if this can lead to a process. In the mean time, very high quality monolayer graphene patches that spontaneously form on the SiC surface are optically identified with AFM, ellipsometry, optical transmission or Raman spectroscopy and then patterned using conventional methods.

**Electronic transport and devices on the C-face**

The enhanced mobilities of C-face graphene is an important consideration for micron scale graphene devices including Hall-bars for metrology as well as high frequency transistors. These larger patterned structures do not rely on quantum confinement but rather exploit the intrinsic two dimensional graphene properties. Hall bar resistance standards are based on the quantum Hall effect that requires high mobility graphene that will exhibit well-developed quantum Hall plateaus at relatively high temperatures (i.e. above cryogenic temperatures[31]) and relatively low magnetic fields (a few Tesla). Terahertz transistors require graphene that has a high mobility at high charge densities (see below). These transistors (envisioned for telecommunications) will be used for amplifiers and oscillators. They need to have significant gain at high frequencies, but large on to off ratios (that are essential for digital applications) are not required.

Figure 5 presents progress on monolayer epitaxial graphene Hall bars on the C-face that show the characteristic half-integer quantum Hall effect as in ref [35]: (a) low charge density allows for QHE plateaus at moderate magnetic field, (b) step free non patterned monolayer flake, (g-f) top-gated monolayer. The high mobility graphene ($\mu \sim 18,000$ cm$^2$V$^{-1}$s$^{-1}$ at a charge density n$\sim 10^{12}$ cm$^{-2}$) is typical for the C-face.

More precisely, the negative charge n$_S \sim$5x 10$^{12}$cm$^{-2}$ at the interface with SiC, can be partially compensated by exposition to the environment, and tuned from *n* to *p*. In Fig 5a, the C-face monolayer was coated with PMMA, rinsed in hot acetone and hot water to remove resist residues. The process was repeated several times until the charge density was adjusted to a value well below n$_S$<1x10$^{12}$cm$^{-2}$. It is interesting that even at this low density (*n*- doping n$_S$= 0.19x 10$^{12}$ cm$^{-2}$) the resistivity $\rho_{xx}$=800$\Omega$ remains low, yielding a Hall mobility

$\mu=1/(n_S e\rho_{xx})=39,800$ cm$^2$/Vs. The Hall resistance $\rho_{xy}$ is linear at low field (away from the QHE plateaus) and does not present the kink or rounding previously associated with the contribution of multilayers[31]. A higher field, the Hall resistance $\rho_{xy}$ shows two very well defined quantum Hall plateaus and the onset of the third one at $\rho_{xy}=(h/\nu e^2)$, with the filling factor $\nu=2$, 6 and 10 ($\nu=4n+2$; $n=0, 1, 2$) expected from monolayer graphene. As already observed on both C- and Si-faces[14, 31] the $\nu=2$ plateau extends in a large field region, over 4.5 Tesla and up to the highest field, 7 Tesla in this experiment. The resistivity $\rho_{xx}$ remains vanishingly small in the plateau region. All these observations corroborate that this is a high mobility low *n*-doped monolayer. Low doping density and high mobility can be finely tuned by environment exposure both for *p*- and *n*-doping. This demonstrates that epitaxial graphene can be used as QHE resistance standards even in moderate magnetic field. Similar results are consistently found on all the monolayer epitaxial graphene devices measured. Another example is given on Fig. 5b for a non patterned monolayer. Here also the $\nu=2$, 6, 10 plateaus are well observed concomitant with zero resistivity at high field. Note also the absence of a sizable localization peak at low field in all devices.

Applying top gate electrodes for individual device switching significantly decreases the mobilities, as already reported[28]. Nevertheless, useful devices can still be made with significantly large mobilities to demonstrate the quantum Hall effect as shown in Fig. 5 d-f. Note that alternative back-gating strategies, which work in a certain temperature range, are useful for global gating in a reduced temperature range[29]. A top gate was deposited on another monolayer by evaporating aluminum metal at a slow rate (0.1 A/s) in low vacuum, allowing aluminum to oxidize during deposition. A subsequent deposition at higher rate (0.9 A/s) covers the alumina with an Aluminum metallic gate. Results for a Hall bar (width 1μm, length 4.3 μm) are presented in Fig. 5d-f. The resistivity as a function of gate voltage at 4K undergoes a maximum for $V_g\sim 0.3$V, clearly showing the conductance modulation by the electrostatic field on both sides of the Dirac point (Fig. 5f) with a high to low resistivity ratio $R(V_g=3V)/R_{max}=I_{off}/I_{on}=13$. This value is common for 2D monolayer graphene. The gate efficiency to change the charge density $n_s$ can be directly measured by the linear Hall resistance $\rho_{xy}$ for different gate voltages at low field. The gate voltage $V_g=0.3$V at the Dirac point gives a charge density $n_S=6.2\times 10^{11}$cm$^{-2}$ induced by the gate to compensate for the natural sample doping. This very close to the doping $n_S=5.6\times 10^{11}$cm$^{-2}$ measured from $\rho_{xy}$ at zero gate voltage.

A gate voltage sweep at 9 Tesla shown in Fig. 5d reveals fully developed quantum hall plateaus at $\sigma_{xy}=\nu e^2/h$ for $\nu=2$, 6, 10, 14 ($\sigma_{xy}$ is the transverse conductivity). The plateaus corresponding to resistivity $\rho_{xx}$ minima and a vanishing resistivity for the $\nu=2$ plateau clearly demonstrates that this is a monolayer. Field sweep at two gate voltages, in the *p*- and *n*-doped region, shows a well-resolved $\nu=2$ plateaus, that extends in a large field region. In particular this shows the QHE effect can be switched from *p* to *n* using the top gate. Note that the mobility for *n* doping is significantly lower than for *p* doping. As for other monolayer graphene of high mobility a very small and narrow weak localization is observed at zero field.

Finally the reciprocal relation between the resistivity and the charge density $n_S = 1/(e\rho_{xy})$ is depicted in Fig. 5c for micron size C-face monolayer graphene Hall bars. As expected for monolayer graphene $\rho_{xx}$ increases at low charge density, but it is interesting to note that $\rho_{xx}$ is quite low compared to results reported for exfoliated graphene, where $\rho_{xx} \approx (4\text{-}6e^2/h)^{-1} = 4\text{-}6\ k\Omega$[36]. The resistivity saturates at high charge density to a value remarkably close to the $1/(80\ e^2/h) = 320\ \Omega$ that was estimated by Orlita et al. [37] from their observation of a scattering time inversely proportional to the energy ($\hbar/\tau = 0.026E$) in IR spectroscopy measurements for the screened inner layers of multilayered epitaxial graphene. The charge inhomogeneity $\delta n_S$ can be estimated[38] from the width of $\rho_{xx}(V_g)$, for top gated samples. From $\rho_{xx} = \rho_s + (e\mu \sqrt{n_S^2 + \delta n_S^2})^{-1}$ we find $\delta n_S = 2 \times 10^{11} cm^{-2}$, $\rho_s = 130\ \Omega$ and $\mu = 8500\ cm^2/Vs$ (*p*-doped side) ($\mu = 6200\ cm^2/Vs$, *n*-doped side) for the sample in Fig. 5f, in very good agreement with the Hall mobility values. The $\delta n_S$ values found for the top gated monolayers are of the same order as monolayers on $SiO_2$. It appears therefore that charge inhomogeneity alone cannot account for the low $\rho_{xx}$ values.

The basic structure and characteristics of a high frequency graphene field effect transistor is shown in Fig. 6. The device has elongated source, drain and gate structures (typically of the order of several microns) in order to ensure high current operation. The other dimensions are optimized for high frequency performance. In particular the channel is ideally short and the contact resistances are ideally low[10,13]. Parasitic capacitances between the various components are designed to be as small as possible. While the highest frequency operation of graphene based transistors has slowing been inching up, the current goal of THz operation has not yet been achieved. However the problems appear to be primarily technical and surmountable[39]. While epitaxial graphene as a viable channel material for THz applications has been established, there are still considerable problems in providing low resistance leads, which determine the access resistance. This may be surprising since from Fig. 5c, the conductivity of (doped) graphene ($\approx 10\mu\Omega cm$) is comparable to that of copper at room temperature. Nevertheless, the minimum resistance of a graphene lead (the so called access resistance) is of the order of a few hundred $\Omega$ per square so that for example, the source lead with a length to width aspect ratio of 10 to 1 has a resistance of several $k\Omega$. This can be reduced by cladding the graphene with a metal, however the metal to graphene resistance is also large. Currently the minimum access resistance reported is about 350 $\Omega$ per micron of channel width[12]. This is still too large for THz applications.

**Structured growth on the Si face.**

Graphene is known to nucleate at step edges and growth rates are known to depend on the crystal face. Growth is considerably slower on the Si face than on the

C face. These facts are advantageously used in the structured growth method (also called the templated growth method[30]) where the silicon carbide surface is first etched to produce steps. During the high temperature annealing using the CCS method, the etched steps crystallize and a graphene film forms on them. The growth times and temperatures are adjusted so that a monolayer graphene film forms only on the step edges and not (or minimally) on the (0001) surfaces.

High temperature annealing causes vertically etched steps (on the order of 10 nm deep) to produce (1-10n) facets with a normal that has an angle of 23 deg with respect to the {0001} direction.

Cross sectional, high resolution transmission electron microscopy by Norimatsu et al.[40] of the graphene on steps on SiC have shown that the graphene terminates perpendicular to the silicon carbide surface both on the bottom of the step or on the steps themselves. This effect has also been observed using scanning tunneling microscopy on small graphene islands on (0001) SiC[27]. Moreover, these atomic resolution studies further show, that the graphene edges are along the zigzag direction, indicating that the graphene sidewall ribbons are zigzag ribbons. This is very important since the edge structure determines the electronic properties of the ribbons. In particular, zigzag ribbons are always metallic[17].

As noted by Kusunoki et al[41], the zigzag nature of the graphene edges is further confirmed in electron diffraction images of carbon nanotubes grown on SiC, which always are of the zigzag variety, which indicates that they terminate with a zigzag edge on the silicon carbide surface. This very important observation, that sidewall graphene ribbons have zigzag edges that terminate in the silicon carbide has profound implications for graphene nano-electronics. For one thing, it implies that the edges are passivated and well defined, that is, they are not subject to the chemical reactivity and the structural disorder of lithographically patterned graphene structures.

The procedure to produce sidewall graphene structures is outlined next (see also Fig. 7). The Si face of 4H (or 6H) SiC is lithographically patterned in using standard lithography methods to produce an etch mask (using, for example, a lithographic resist such as PMMA) on the surface. The masked surface is then subjected to a plasma etch (typically with $SF_6$) to etch the desired pattern to a predetermined depth into the surface. This procedure can be repeated several times to produce structures with varying depths. Alternatively, wide ribbons can be produced by fusing narrower ribbons as shown in Fig. 7b. The resulting structure is then annealed using the CCS method[23] at a temperature of about 1550 C for about 10 minutes, which anneals the steps (typically producing on facets where n ranges from 1 for shallow trenches to about 10 for deep ones) with typically a monolayer of graphene on the steps. Next, if desired, gate structures can be patterned on top of the sidewall ribbons in the usual way. It is occasionally desirable to etch away spurious graphene ribbons that have formed on the sidewalls of steps on the substrate that are inevitably present due to the slight miscut of the original crystal

(Fig. 7a). However, judicious choice of the orientation of the patterned structure with respect to the miscut, minimizes this problem.

Figure 7 shows several examples of several sidewall graphene structures produced by this method, to demonstrate its effectiveness. Atomic force microscopy (AFM) images present the topography, while electrostatic force images (EFM) provide a characteristic contrast that distinguishes graphene from silicon carbide (Fig. 7).

Narrow graphene ribbons are particularly important for graphene-based nanoelectronics. As mentioned above, sidewall ribbons are expected to have zigzag edges, which we find are (always) metallic. It should be noted that, on the contrary, all graphene ribbons produced to date using standard lithography methods are found to be semiconducting, i.e. have a transport gap a low temperature or close to the charge neutrality point. This is actually not expected since graphene ribbons are generally expected to be metallic[17] (the only exception being specific graphene ribbons with armchair edges mentioned above). It is now generally accepted that the bandgaps observed in narrow graphene ribbons produced by standard lithography methods are in fact mobility gaps caused by the rough edges[6-8, 19]. In contrast all of the gated sidewall graphene ribbons produced by the methods outlined here are metallic, consistent with expectations. Moreover, we have observed that a majority of these sidewall ribbons show evidence of (quasi) ballistic transport involving a single conducting channel with mean free paths on the order of 1 μm, which is significantly greater than the typical mean free path in 2D graphene. The narrow graphene ribbons produced for these studies are produced as follows (see Fig. 8).

Two opposing deep (>100 nm) U shaped trenches are first etched. Subsequently a narrow (~10 nm) trench is etched, connecting the U trenches (Fig 8 a). After CCS thermal annealing, the sidewalls of the U trenches form the broad graphene leads for the two narrow graphene sidewall ribbons between the U's. Metal contacts provided to the graphene leads allow for 4-point transport measurements of the graphene ribbon. Figure 8 provides some details of the production and characterization methods used. Specifically, as shown, the EFM resolution of is 30 nm, which is considerably greater than that of Raman (with about 1μm resolution) for detecting graphene nanostructures. Moreover the EFM method can be improved, by modeling the tip to graphene interaction, to provide a resolution of 10 nm.

While structured graphene is most suited for graphene nanostructures, it can also be used for Hall bars and other micron sized devices. An example is shown in Fig. 9 where a graphene ribbon that is about one micron wide and several microns long is connected to 6 leads in the Hall bar configuration for transport measurements. For this sample, a distinct anomalous diode like effect is observed in this ribbon when the resistance is measured between two opposing electrodes as shown in Fig. 9c. Ideally, the resistance (in the absence of a magnetic field) is expected to be zero, however a significant resistance is obtained. This normally indicates a misalignment of the voltage contacts. However the diode-like behavior signifies that nonlinear effects are playing a role. Note that at 4 K, the resistance even changes sign. Similar non-linear properties are seen in nanoscopic phase coherent ballistic junctions and it is expected that coherent transport plays a role in this case as well. This interpretation is fortified by the observation that the anomalous behavior vanishes at room temperature, which is consistent with expected reduction in the phase coherence length (to < 20 nm) at room temperatures. As expected, this wide side- wall Hall bar presents regular Shubnikov de Haas oscillations, like for other micron size patterned ribbons[4] (fig. 9d). The position of the resistance maxima in field (Landau plot) is a straight line from which the carrier density is determined to be $n_s$= $5.1 \times 10^{12}$cm$^{-2}$. The magnetic field was tilted at 23 deg from the normal of the SiC (0001) surface to be approximately perpendicular to the sidewall ribbon.

Patterned sidewall rings connected to wide ribbons that serve as contacts probes show distinct magnetoresistance oscillations, as shown in Fig.9 e. The Fourier transform (Fig. 9f ) reveals the regularity of the oscillations. The arrows indicate the positions where Aharonov-Bohm quantum interference maxima (fundamental and overtones) are expected. This clearly indicates that phase coherence length is of the order of the size of the ring.

## Conclusion

We have presented recent results on a variety of patterned graphene structures, on both the carbon and silicon terminated faces of hexagonal silicon carbide. We have demonstrated that structured graphene growth is a powerful and flexible method to control the shapes of the graphene structures and that a wide variety of structures can be produced both for fundamental studies as well as for applications. It presents an important step towards the realization of high mobility quasi one dimensional graphene structures that do not suffer from the strong localization effects observed in conventionally patterned graphene structures, that all but obviate graphene's advantages at the nanoscale.

## Acknowledgements

This research was supported by the W. M. Keck Foundation, the Partner University Fund from the Embassy of France, the AFSOR grant No FA 9550-10-1-0367 and the NSF MRSEC Program under Grant No. DMR-0820382. We thank E. Conrad, M. Kindermann and Z . G . Jiang for insightful discussions.

(continued) graphite: 2D electron gas properties and a route toward graphene-based nanoelectronics. J Phys Chem B 2004, 108, 19912.

**FIGURE CAPTIONS**

**FIG 1**. AFM and optical images of graphene the 000-1 face of 4H-SiC . (a) AFM image of a graphene monolayer on a stepped 000-1 surface; the graphene layer exhibits pleats (bright lines) and is confined to the central patch. Note the molted appearance of the areas not covered by graphene. Bright spots are due to post-production contamination; (b) AFM image of a monolayer graphene patch draped over the distinct central stepped region of the image that extends to the top. (c) Optical image of monolayer graphene growth around etched trenches in the SiC surface (that resulted from random cracks in the polymer etch mask). The graphene

stands out as slightly brighter patches decorating the dark trenches. (d) Mulilayer graphene growth at a screw dislocation.

FIG 2. (a) Typical examples of micro Raman spectra used to identify monolayer graphene, black trace: graphene, red trace, bare SiC substrate (Raman laser wavelength λ =532nm, spectral resolution Δ =0.3cm$^{-1}$). Note the G band at 1585 cm$^{-1}$ and the prominent 2D band at 2685 cm$^{-1}$ that can be fit with a single Lorentzian (with a typical width of 30 cm$^{-1}$). (b) peak position of the G band; (c) peak position of the 2D band as a function of charge density.

FIG 3. Detailed analysis of a monolayer graphene patch on the C-face. (a) AFM topology map (Scale bar = 10 µm), showing the monolayer patch draped over several substrate steps. The graphene layer can be recognized by the pleats (white lines) that crisscross the patch. (b) Raman intensity map of the 2D Raman band recorded at 2678 cm$^{-1}$ clearly revealing the graphene covered region (same scale as (a)). (c) Raman intensity map recorded at 2764 cm$^{-1}$ showing that this anomalous pleat-free region is covered with a monolayer of strained graphene. (d) Micro Raman spectra of areas in b (black) and c (red) showing a considerable shift in the later, that may be caused by an anomalously strong coupling to the substrate

FIG 4. Monolayer graphene growth on 4H-SiC (000-1) on a pre-patterned structure indicates preferred sites for graphene growth. (a) AFM image for hexagon-trench structure after graphitization. Structure is etched to a depth of∼50nm using the Inductively Coupled Plasma (ICP) method. SiC step flow inside the trench. Faint graphene pleats are visible. (b) EFM amplitude scan shows contrast between darker graphene regions and lighter SiC regions as verified in the Raman 2D band intensity map (c). Scale bar: 2µm.

FIG 5. The quantum Hall effect in epitaxial graphene C- face for micron wide samples. (a) Hall bar (µ=39,800 cm$^2$V$^{-1}$s$^{-1}$, n= 0.19x10$^{12}$ cm$^{-2}$) ; (b) monolayer step free non patterned flake (ρ$_{xx}$=320 Ω, µ=21,100 cm$^2$/Vs at charge density n$_S$= 0.92 x 10$^{12}$cm$^{-2}$) ; (c) Plot of resistivity versus charge density for micrometer size monolayer C-face Hall bars. The dotted line correspond to ρ**=**(80e$^2$/h)$^{-1}$ from ref.[42]; (d-f) top gated Hall bar (1µmx 4.3µm). (d) Resistivity ρ$_{xx}$ (black) and Hall conductance σ$_{xy}$ (red) versus gate voltage at 4K and 9 T, showing plateaus in the *n*- and *p*-doped regimes; (e) Hall resistance ρ$_{xy}$ versus magnetic field measured at two gate voltages and comparable n and p charge density; red curve: V$_g$=0.4V, *n*-doped region (n$_s$=5x10$^{11}$ cm$^{-2}$, µ=6,200cm$^2$/Vs), black curve: V$_g$=0V, p-doping (n$_s$=5.6x10$^{11}$ cm$^{-2}$, µ=8,500cm$^2$/Vs). A significant reduction in the mobility (compared with bare samples above) is caused by the application of the top gate; (f) Resistivity ρ$_{xx}$ as a function of gate voltage at 4K and zero magnetic field; Inset: the measured Hall bar.

**FIG 6.** C-face graphene FET transistor for high frequency applications. (a) Source Drain current versus Source Drain voltage. Maximum current density > 1.5 mA/μm; $g_m$ > 0.4mS/ μm; n = $1.6 \times 10^{12}$ /cm$^2$; $\mu_{FET}$ = 8700 cm$^2$/Vs. The various curves (from bottom to top) have been obtained for Vg from 1.5 V to -1.5 V in 0.5 V increments. (b) The gain ($H_{21}$) versus frequency measured up to 50 GHz. The extrapolated unit gain cutoff frequency $f_T$= 90 GHz (after de-embedding). (Inset) Optical microscopy image of the transistor (gate length is 150 nm). It is expected that $f_T$ will increase significantly by reducing the gate length to 15 nm.

**FIG 7.** Examples of sidewall graphene structures, showing atomic force microscope (AFM) images (top row) and electrostatic force microscope (EFM) images (bottom row). Top and bottom rows are at the same scale. The EFM images provide a bright contrast for graphene and a dark contrast for SiC, thereby providing a simple method to locate graphene covered surfaces.(a) Sidewall graphene ribbons grown along bunched SiC step edges of steps due to the slight miscut of the crystal. (b) Sidewall graphene growth along patterned SiC surface that was etched to a depth of 15 nm. Note the serpentine sidewall that was etched to provide for a wide lead (after annealing) for the narrow, straight ribbon segments, demonstrating the flexibility of the method.  (c) Graphene ring supplied with leads. The original patterned SiC structure consisted of a circular SiC mesa connected to radial SiC walls. Graphene grown on the sidewall of the mesa produced the graphene ring and graphene grown on the walls produced the leads. HSQ was used as etching mask. (d) Array of graphene nanoribbons. Narrow parallel groves were patterned covering the entire SiC chip. More than 30% of the surface is covered with graphene ribbons. The SiC etching depth was 150nm and photolithography was used to produce the Ni etching mask.

**FIG 8**
Examples of structured graphene nanoribbon and connection for 4-probe measurement (a) AFM image of U-shaped deep trenches connecting a shallow trench overlapped with EFM of a Hall bar structure. (b-d) zoom on the shallow trench (b) AFM image after SiC patterning, (c) AFM image after subsequent graphitization, showing the rounding of the trench due to SiC step flow at high temperature (d) EFM image of (c) showing two nanoribbons grown on the side wall of the trench. From ref[30].

**FIG 9**
Examples of structured graphene and their electronic properties (a) 3D AFM topology image overlapped with EFM of a Hall bar structure. Bright area on the sidewalls shows where graphene has formed.  (b) is a SEM image of a Hall in (a) after contact pads  (light grey squares) were patterned. (c) $V_{CD}$ versus $I_{AB}$ measurements at various temperatures of the structure in (b) showing ohmic behavior at room temperature and non-linear behavior at low temperature.  (d)

Shubnikov de Haas oscillations observed of the sidewall Hall bar shown in the lower inset. The magnetic field is at 23 deg from the normal of the SiC (0001) surface and approximately perpendicular to the sidewall ribbon. (Upper inset): Landau plot of the peak positions from which the carrier density is determined to be $5.1 \times 10^{12}$ cm$^{-2}$. (e) Magnetoresistance oscillations of the 7.1 µm diameter sidewall ring structure shown in (f) (inset). (f) Fourier transform of (e) showing distinct maxima. The arrows indicate the positions where Aharonov-Bohm quantum interference maxima (fundamental and overtones) are expected.

Fig.1

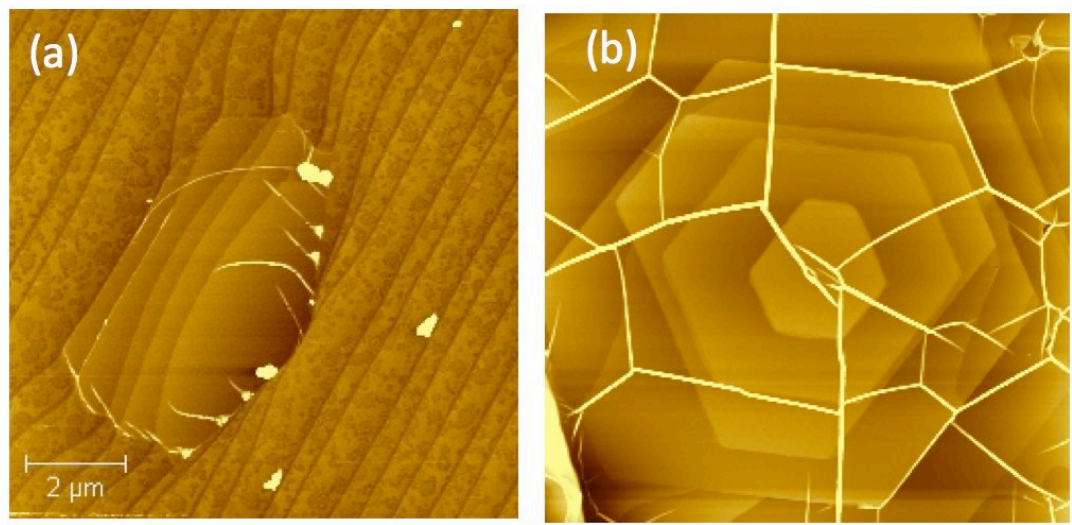

Fig.2

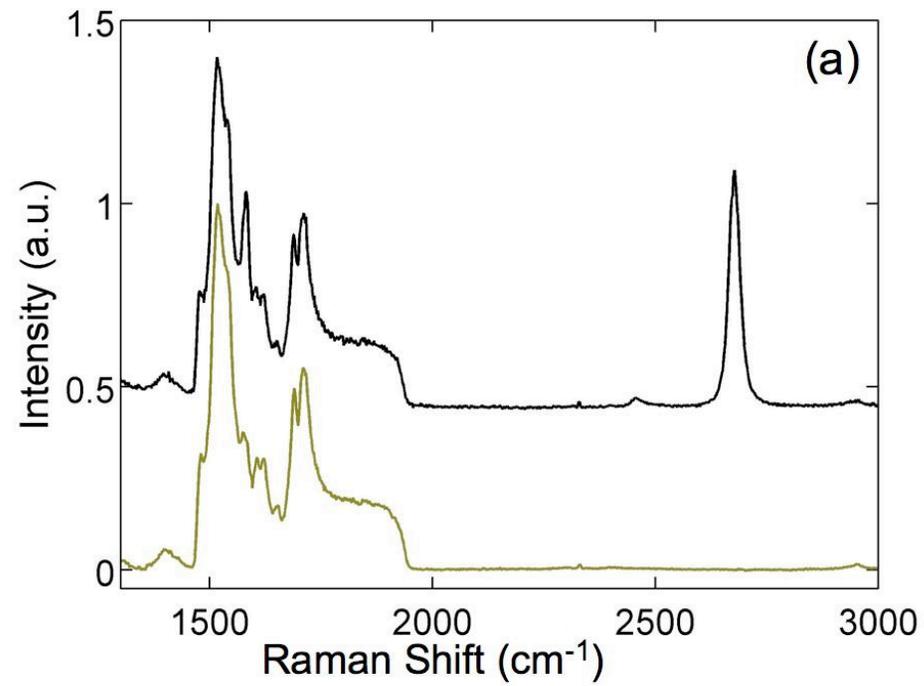
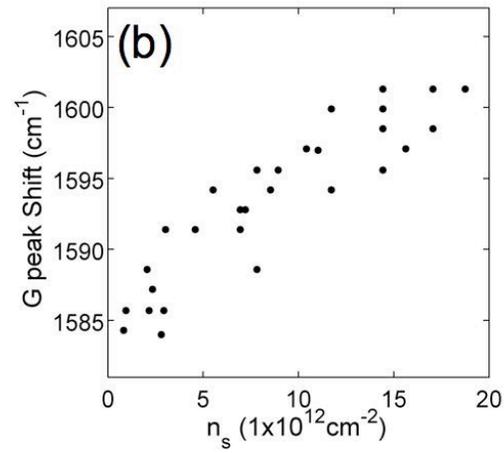
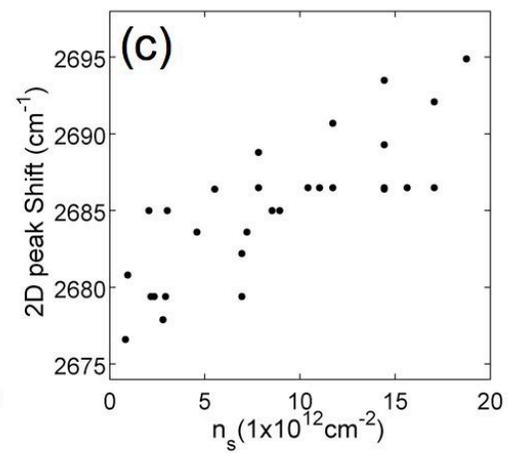

Fig.3

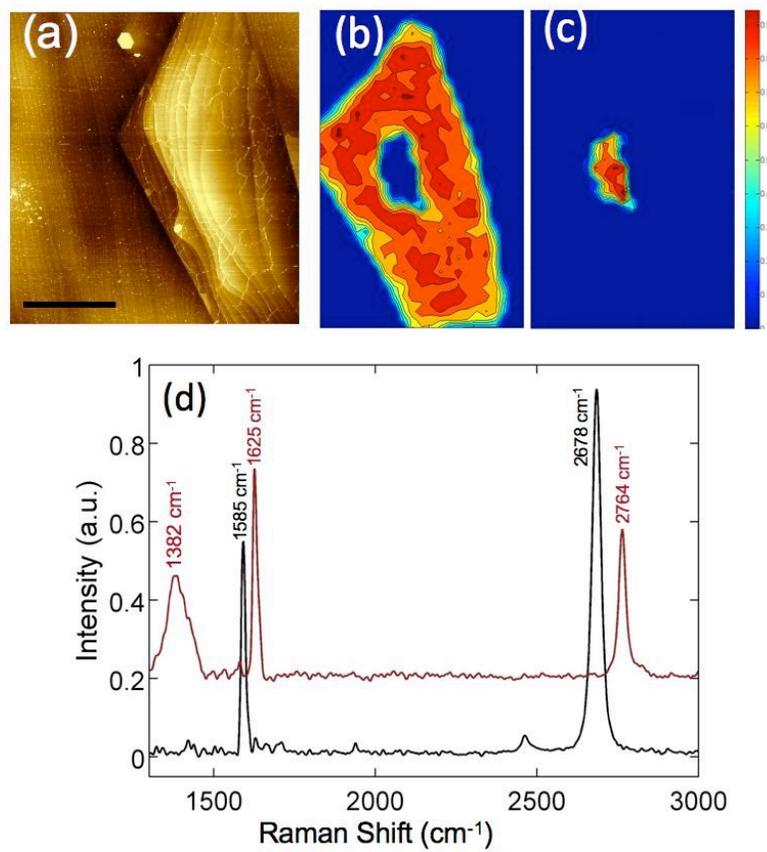

Fig.4

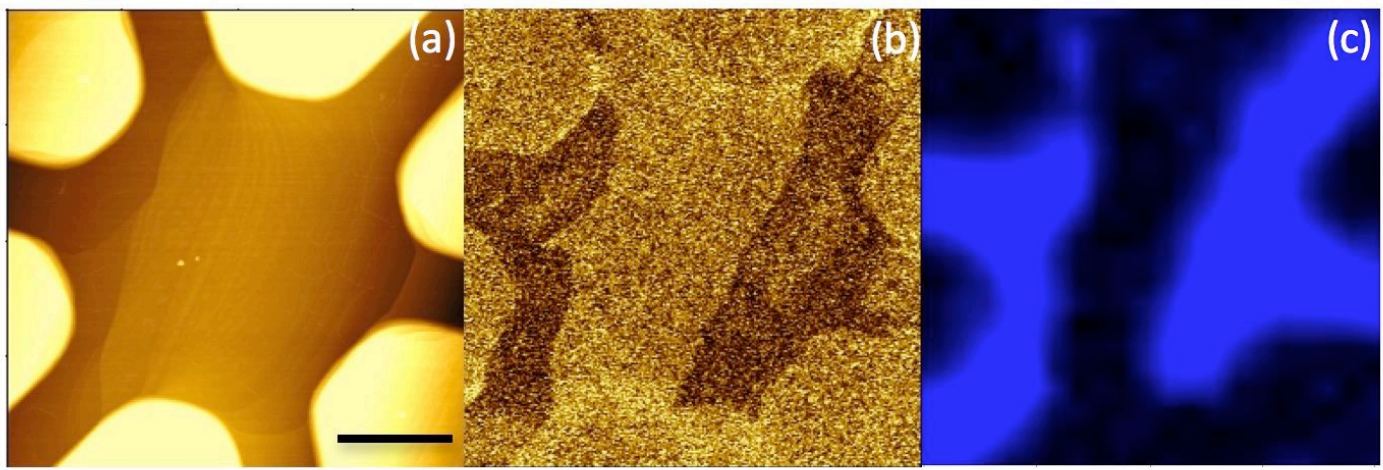

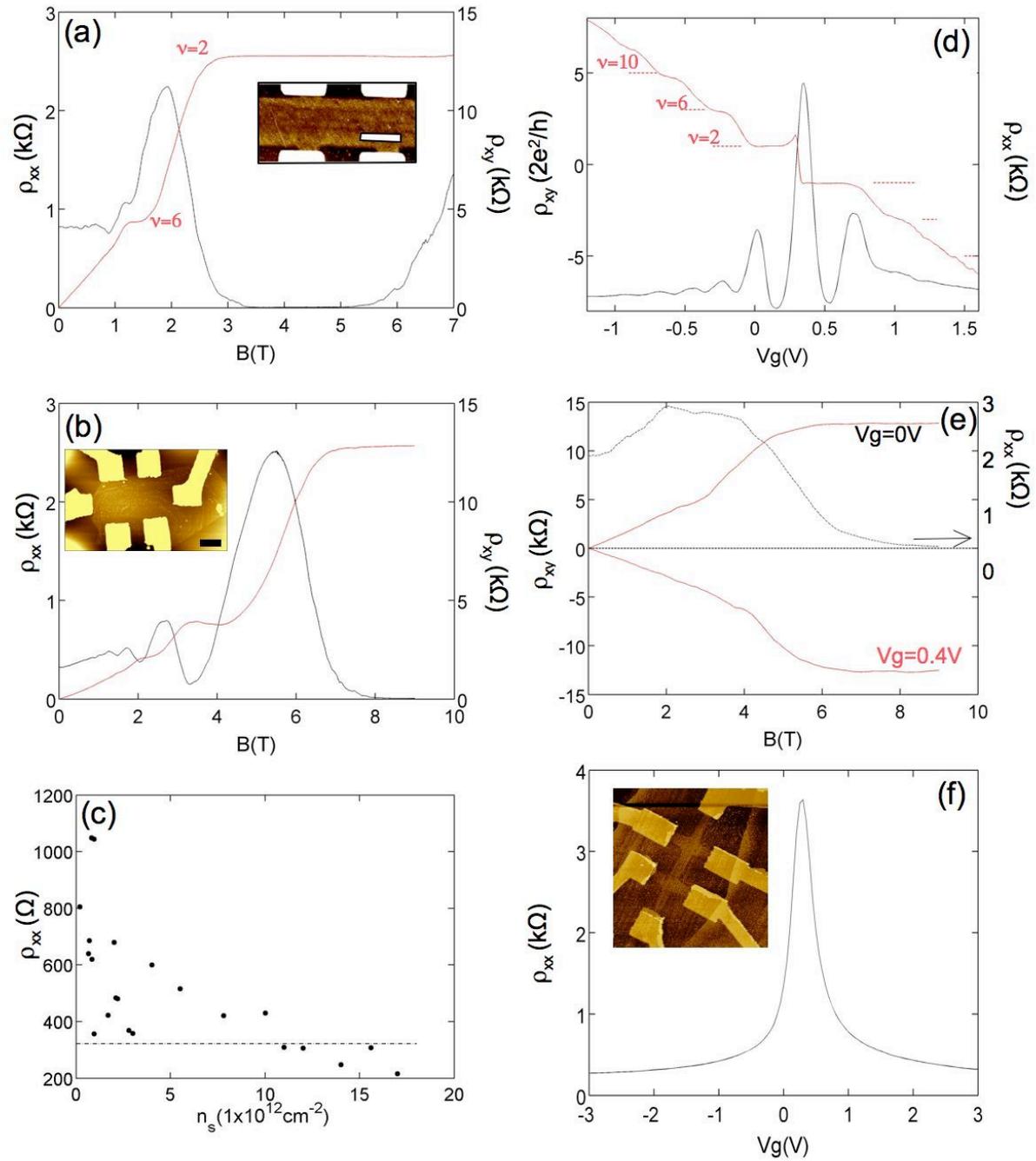

Fig.5

Fig.6

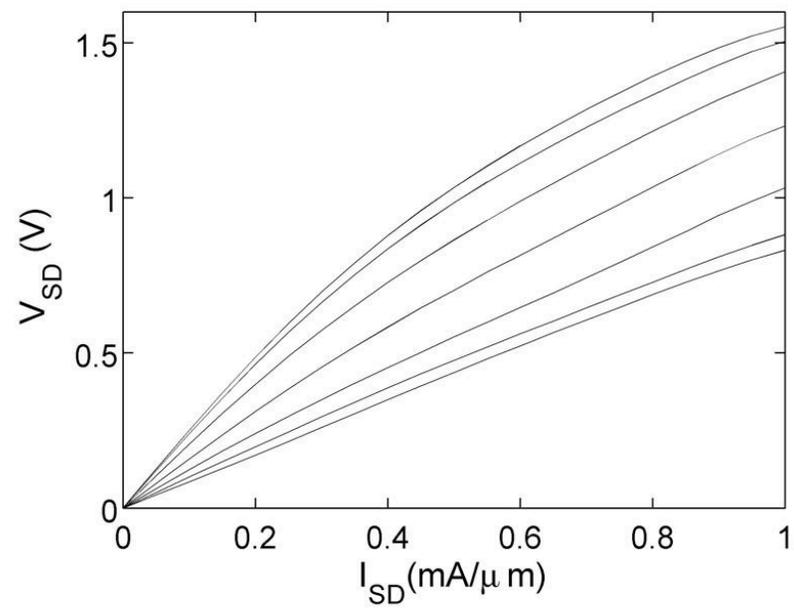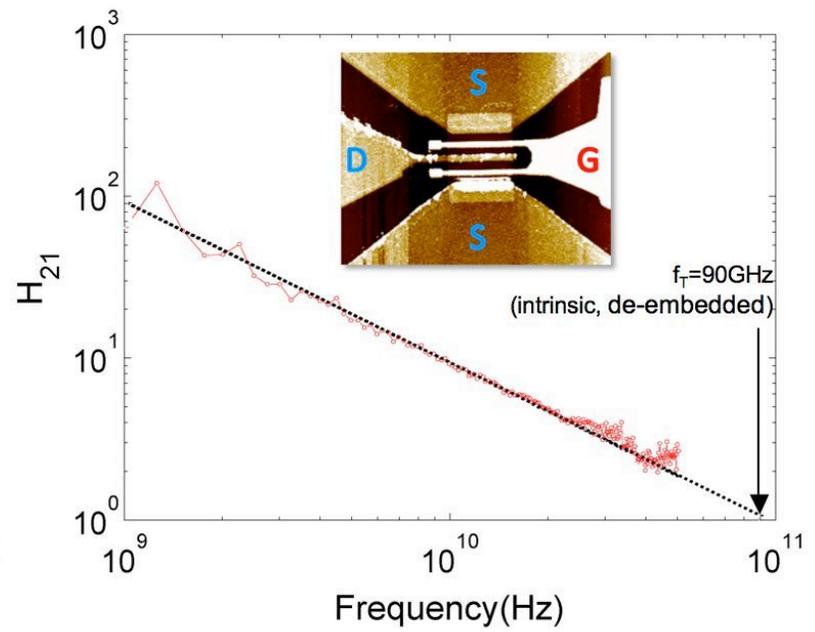

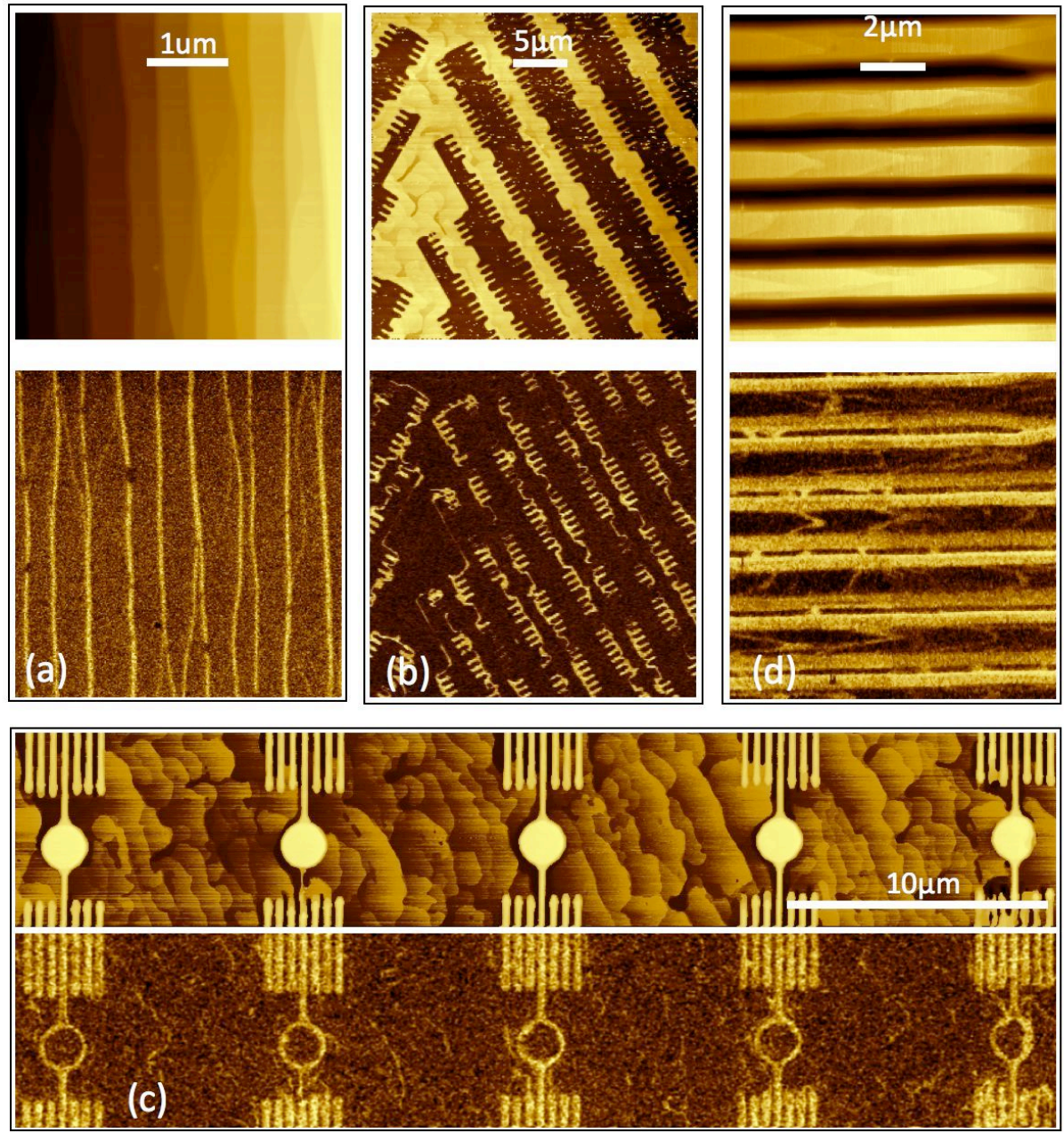

Fig.8

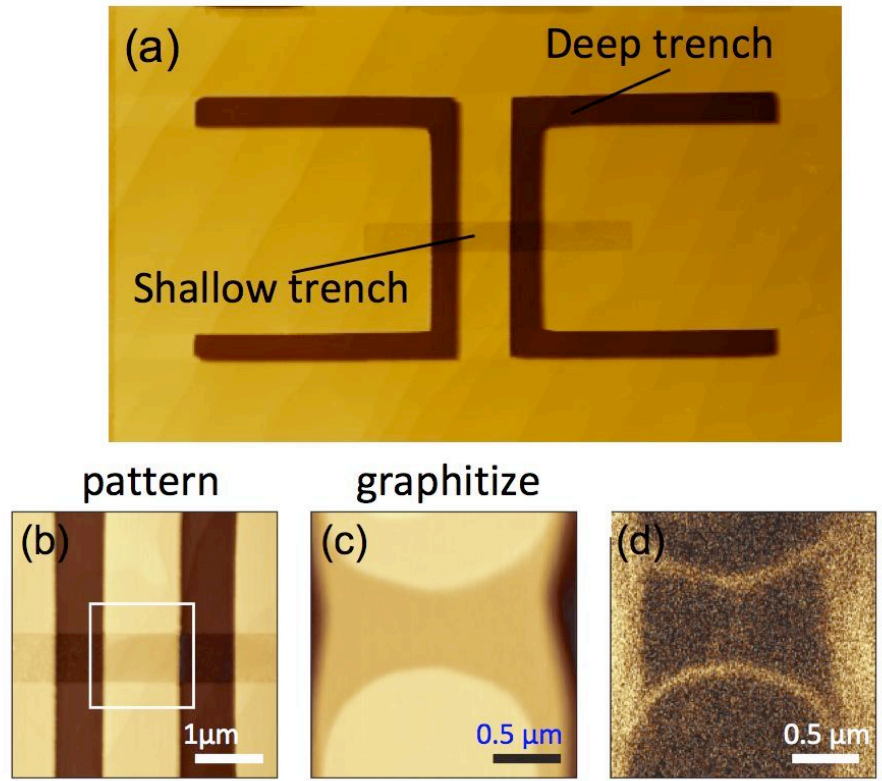

Fig.9

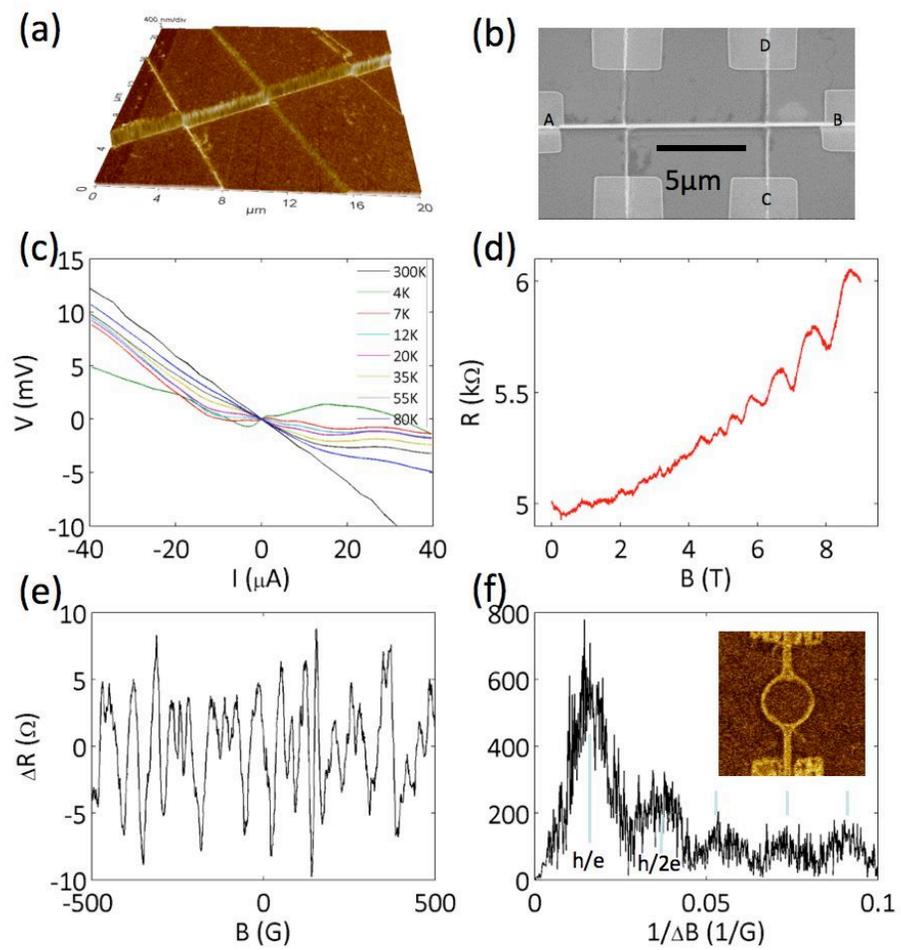